\documentclass[10pt,twocolumn,letterpaper]{article}

\usepackage[pagenumbers]{cvpr} 

\usepackage[dvipsnames]{xcolor}

\usepackage{algorithm}
\usepackage{algorithmic}

\usepackage{threeparttable} 
\usepackage{amsmath,amsfonts}
\usepackage{multirow}

\definecolor{cvprblue}{rgb}{0.21,0.49,0.74}
\usepackage[pagebackref,breaklinks,colorlinks,citecolor=cvprblue]{hyperref}

\def\paperID{00000} 
\def\confName{CVPR}
\def\confYear{2024}

\title{RobWE: Robust Watermark Embedding for Personalized \\ Federated Learning Model Ownership Protection}

\author{Yang Xu$^1$,Yunlin Tan$^1$, Cheng Zhang$^1$, Kai Chi$^1$, Peng Sun$^1$, Wenyuan Yang$^2$, \\Ju Ren$^3$, Hongbo Jiang$^1$, Yaoxue Zhang$^3$\\
	$^1$Hunan University, $^2$Sun Yat-sen University,$^3$Tsinghua University\\}

\begin{document}
\maketitle
\begin{abstract}
 Embedding watermarks into models has been widely used to protect model ownership in federated learning (FL). 
 However, existing methods are inadequate for protecting the ownership of personalized models acquired by clients in personalized FL (PFL). 
 This is due to the aggregation of the global model in PFL, resulting in conflicts over clients' private watermarks. Moreover, malicious clients may tamper with embedded watermarks to facilitate model leakage and evade accountability.
 This paper presents a robust watermark embedding scheme, named RobWE, to protect the ownership of personalized models in PFL. 
We first decouple the watermark embedding of personalized models into two parts: head layer embedding and representation layer embedding.
 The head layer belongs to clients' private part without participating in model aggregation, while the representation layer is the shared part for aggregation.
 For representation layer embedding, we employ a watermark slice embedding operation, which avoids watermark embedding conflicts. Furthermore, we design a malicious watermark detection scheme enabling the server to verify the correctness of watermarks before aggregating local models.
 We conduct an exhaustive experimental evaluation of RobWE.
 The results demonstrate that RobWE significantly outperforms the state-of-the-art watermark embedding schemes in FL  in terms of fidelity, reliability, and robustness.
\end{abstract}
\section{Introduction}
\label{sec:Introduction}
Training artificial intelligence models  demands substantial resources, and thus well-trained commercial models have evolved into pivotal assets for many corporations~\cite{he2022cater, skylar_exploring_2023,yue_learning_2023}. 
Unfortunately, models are confronted by pronounced risks of intellectual property infringement.
Previous studies have  unveiled that models may be illicitly leaked during usage or even stolen through model imitation attacks~\cite{197128, Krishna2020Thieves, Zhang_thesecret_2020, shen_model_2022}.
Therefore, guaranteeing ownership of models has become an urgent imperative.

Model watermarking is an effective technology for model ownership confirmation. 
Specifically, model watermarking involves embedding identifiable or distinctive markers into models based on features or backdoors.
The model owner can proof the ownership by extracting the watermark in the model when the ownership dispute occurs.
Initial model watermarking studies primarily focused on guaranteeing ownership in traditional machine learning scenarios.
Recently, with the rise of federated learning (FL), there has been a considerable surge of interest in protecting model ownership within distributed machine learning scenarios.
FL, as a privacy-preserving machine learning, facilitates collaborative training of models without collecting client data~\cite{pmlr-v54-mcmahan17a}.
However, as each client in FL has access to the trained models, FL is susceptible to heightened risks of model leakage.
Besides, the distributed model training process of FL also poses challenges for watermark embedding and identification.

To protect the intellectual property of models in FL, federated model ownership verification schemes have been proposed.
Waffle ~\cite{tekgul_waffle_2021} is the pioneering effort. 
It proposed to embed watermarks when the server aggregates local models to protect global model ownership.
Considering client contributions within FL, subsequent approaches ~\cite{liu_secure_2021, li_fedipr_2022} permit clients to embed private watermarks during local model training, facilitating independent verification of model ownership by all parties.
However, these schemes confine to the vanilla FL paradigm, rendering them susceptible to performance degradation in real-world scenarios where the data among clients is non-independent and identically distributed (non-IID).

To account for such data heterogeneity, researchers have proposed the framework of personalized federated learning (PFL), such as clustered FL ~\cite{ghosh2020efficient, sattler_clustered_2021, donahue_model-sharing_2021} and federated meta learning \cite{fallah2020personalized,chen_metafed_2023}. 
PFL not only leverages the collective knowledge from the distributed data sources to train a shared global model but also allow client to tailor personalized models according to their specific data distributions. 
Unfortunately, while it allows for more tailored and effective models for clients with heterogeneous data distributions or unique preferences, PFL faces the following three-fold challenges in preserving the ownership of these personalized models: 
\begin{itemize}
    \item First, personalized model ownership currently has no corresponding solution. Existing solutions focus on embed client watermarks within the shared global model in FL. If PFL adopts these schemes directly, after aggregation, personalized models will encompass watermarks from other clients, failing to satisfy the private nature of personalized model ownership.
    \item Second, the model obtained by PFL also contains all clients' knowledge contributions. To declare the contributions of all clients, embedding a public watermark into the model is a natural idea. However, watermarks from different clients conflict in the same embedding region. While some preliminary solutions exist, their efficacy is insufficient when the number of participants is large and the watermark bits is long \cite{liu_secure_2021, li_fedipr_2022, Yang2023FedZKPFM}.
    \item Third, some malicious clients do not comply with the watermark embedding protocol.
Instead, they may insert falsified watermarks, making it challenging to verify model ownership when disputes happen.
\end{itemize}

Motivated by the above discussions, this paper presents a robust watermark embedding scheme, named RobWE, to protect the ownership of personalized models in PFL.
RobWE disentangles the watermark embedding process into two distinct components tailored for personalized FL scenarios: head layer embedding and representation layer embedding.
The head layer, constituting the client's private segment, remains separate from model aggregation, while the representation layer is part of the client's shared segment. 
Hence, the watermark slice within the representation layer signifies the client's contribution to the global model, while the private watermark within the head layer elucidates ownership of the personalized model.
Furthermore, we employ a watermark slice embedding operation within the representation layer to mitigate conflicts arising from model aggregation and the embedding of multiple watermarks. This operation also prevents malicious clients from tampering with the unique watermark to evade detection.
Finally, we propose a tampered watermark detecting mechanism to thwart adversaries, which enables the server to validate the accuracy of client watermark slices before aggregating corresponding representation models.
Specifically, the server extracts the client watermark slice from the representation model and determines whether it is honestly embedded by assessing its similarity with watermark slices from other clients with the same embedding frequency. The major contributions of this paper are as follows.
\begin{itemize}
\item To our knowledge, we propose the first robust model ownership protection framework for personalized federated learning. This framework enables private watermark embedding and model ownership verification for clients.

\item By decoupling the watermark embedding process and adopting watermark slices, RobWE effectively addresses the watermark interference issue arising from model aggregation in FL. 
This strategy greatly enhances flexibility, allowing more clients to embed multiple watermarks. 

\item By employing the data distribution characteristics of watermark slices from clients with identical embedding frequencies, we design a tamper-resistant watermark detection mechanism, which can accurately detect malicious clients. 


\item We comprehensively evaluate RobWE and demonstrate that RobWE significantly outperforms the state-of-the-art watermark embedding schemes in FL regarding fidelity, reliability, and robustness. 


\end{itemize}
\section{Related Work}
\label{sec:Related Work}

\subsection{Watermarking in Centralized Learning}
Watermarking is a technique designed to prevent models from being stolen and copied by others. By embedding watermarks into the model, the model owner is able to prove the attribution of the model. One way to achieve this is to change the parameters or the structure of the model. This involves schemes such as embedding bit strings in the loss function~\cite{uchida_embedding_2017} or a particular activation layer~\cite{bita_embedding_2019}, fine-tuning pre-trained models to embed identifiable fingerprints~\cite{chen_deepmarks_2019}, adjusting the GAN-like structure for more covert watermark generation~\cite{wang_riga_2021}, adding a special task-agnostic barrier~\cite{zhang2022Deep} and introducing a passport layer to prevent ambiguity attacks~\cite{fan_deepip_2021, fan_rethinking_2019}. Another way is to use trigger sets to train as a backdoor to the model. Lin~\etal~\cite{Lin2023Triggerset} generated trigger sets using a key from the authority to create a scrambling sequence that shuffles pixels and assigns original labels. Zhang~\etal~\cite{zhang_protecting_2018} introduced various methods to generate watermarking keys, which are then utilized in fine-tuning pre-trained models. Guo and Potkonjak~\cite{guo_evolutionary_2019} contributed a technique leveraging evolutionary algorithms to generate and optimize patterns for backdoor watermarking.

\subsection{Watermarking in Federated Learning} 
To protect the intellectual property of models in FL, 
Tekgul~\etal~\cite{tekgul_waffle_2021} treated the server as the model owner and embed the backdoor-based watermark in the aggregation phase. 
However, the server does not always take ownership of the model. Liu~\etal~\cite{liu_secure_2021} proposed embedding backdoor-based watermarks on the client side.  
Li~\etal~\cite{li_fedipr_2022} proposed FedIPR to embed feature-based and backdoor-based watermarks on all clients. 
This scheme applies to the existing FL problem settings. However, when the number of clients increases, the watermark protection undergoes significant performance degradation.
More importantly, the aforementioned schemes have not accounted for the need for intellectual property protection in PFL scenarios. 
Furthermore, these schemes are susceptible to embedded watermarking attacks, such as model fine-tuning~\cite{uchida_embedding_2017} and model pruning~\cite{bita_embedding_2019}.    
To thwart these attacks, some schemes~\cite{Shao2022FedTracker,Liang2023FedCIP, Yang2023FedZKPFM} were proposed to trace the leaked malicious client back.
Nevertheless, none of them can prevent malicious clients from tampering with the watermark during the watermark embedding phase, let alone the watermark conflicts during model aggregation.  


\section{Problem Statement}
\label{sec:Problem Statement}
In this section, we first present tampering and adaptive tampering attacks during watermark embedding. Then, we formally define the watermark embedding problem in PFL. 

\subsection{Tampering Attack}
In previous studies~\cite{Yang2023FedZKPFM,yang_fedsov_2023}, the server sends a shared watermark to clients, who can maintain ownership of the model by embedding this shared watermark. However, malicious clients can remove the shared watermark by embedding a tampered watermark, rendering it impossible to determine ownership of the leaked model.
Suppose the original watermark is $b_i$, and the watermark tampering rate is $f_{t}$. Launching the tampering attack means flipping the number of bits in $b_i$ by a proportion of $f_{t}$.
Obviously, as $f_{t}$ increases, the detection rate of the original watermark will gradually decrease and even drop to $0$, thus achieving the purpose of model stealing. 


\subsection{Adaptive Tampering Attack}
\begin{figure}[t]
   \centering
   \includegraphics[width=\linewidth]{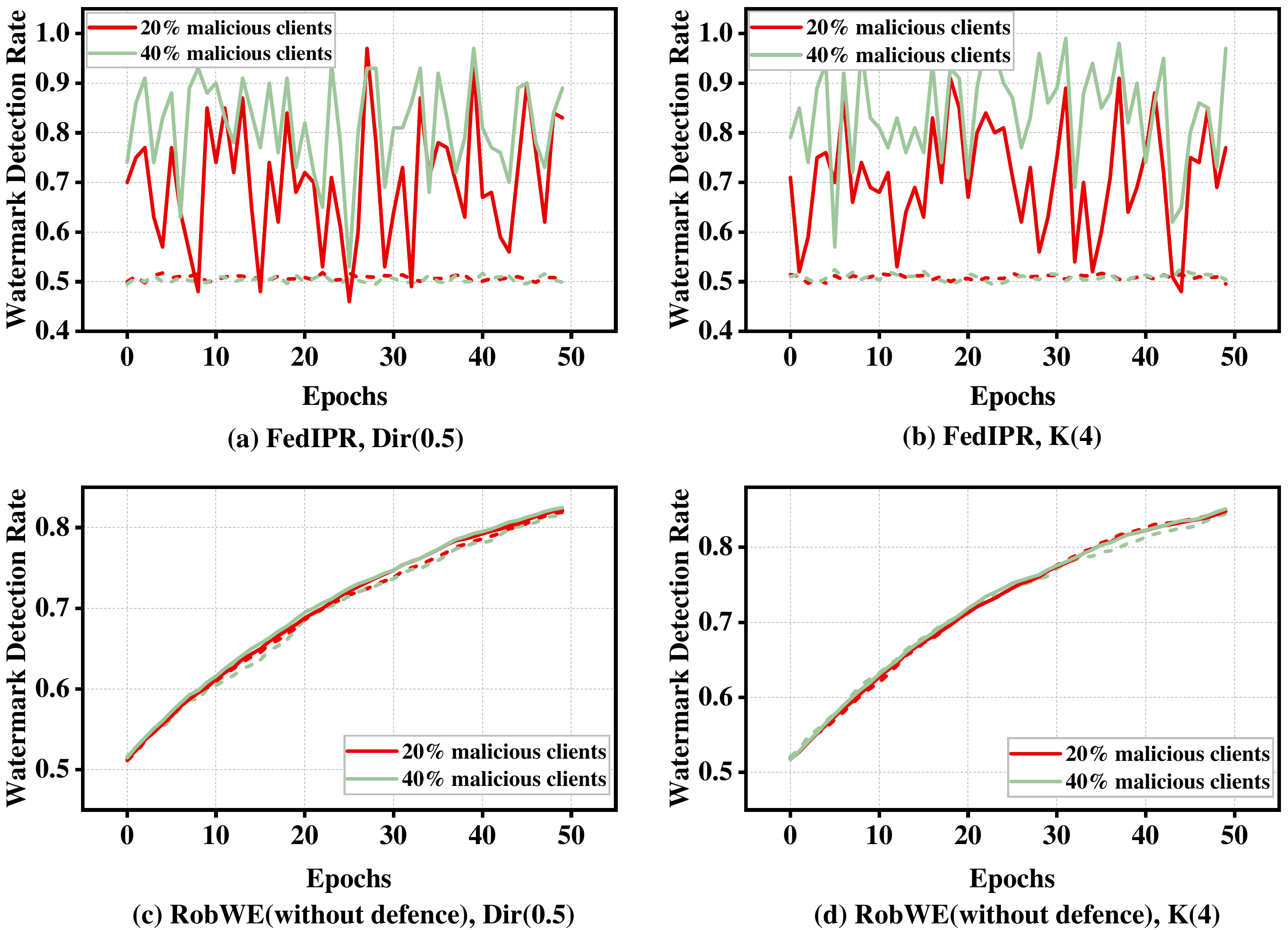}\\
  \caption{In the Non-IID settings (i.e., Dir(0.5) and K(4)), the watermark detection rates of malicious clients (solid line) and honest clients (dashed line) in FedIPR and RobWE (without defense).
  }
  \label{figs:tampered-attack}
\end{figure}

We further consider the evolved version of the tampering attack, where clients can collude to launch tampering attacks, referred to as adaptive tampering attacks. 
And we assume that honest clients are the majority.
Here, we evaluate the effectiveness of the attack against FedIPR~\cite{li_fedipr_2022} and our proposed RobWE. 
In this experiment, we consider $20\%$ and $40\%$ of clients are malicious, who tamper with the watermark to be embedded during the training process, with each attack tampering with $10\%$ bits of the watermark.
We implement our scheme without defense to demonstrate the impact of adaptive tampering attacks.
As shown in~\cref{figs:tampered-attack}(a) and (b), as the ratio of malicious clients launching adaptive tampering attacks increases in FedIPR, the detection rate of tampered watermarks embedded by malicious clients becomes greater than that of the watermarks embedded by honest clients, resulting in the theft of model ownership during the embedding process. 
In comparison, as shown in~\cref{figs:tampered-attack}(c) and (d), under RobWE, without taking any attack detection, malicious clients can only approach the watermark detection rate of honest clients and cannot completely dominate. 

The reason is that RobWE adopts a watermark slice operation, which allows the server to assign different watermark slices to each client and embed them in different regions. In this way, clients are unaware of the other watermark slices, preventing malicious clients from simply tampering with the watermark to remove it.
In contrast, FedIPR embeds the client watermarks in the same region, resulting in malicious clients launching adaptive tampering attacks very easily.
Moreover, watermarks embedded in the same region interfere with each other, which not only poses the capacity issue but also fails to satisfy the urgent demand of PFL for unique private ownership of models.

\subsection{Problem Definition}
Inspired by prior works~\cite{collinsExploitingSharedRepresentations2021,uchida_embedding_2017} and the above observations, we characterize the watermark embedding problem in PFL based on the following two tasks:  
\begin{itemize}[leftmargin=*]
\item \textbf{Main task} is to achieve the goal of PFL.
Here, we adopt the idea of representation learning~\cite{yue_layout_2023,chen_fair_2023, collinsExploitingSharedRepresentations2021} to achieve model training.  
Formally, we consider a PFL setting consisting of a global common representation $m_{r}$, which is a function parameterized by $r\in \mathcal{R}$ learned by all clients, and client-specific heads $m_{h_{i}}$, which are functions parameterized by $h_{i}\in \mathcal{H}$ learned individually by each client $i$.
Here, $\mathcal{R}$ and $\mathcal{H}$ are global common representation solution space and head solution space, respectively.
Then, the global objective of PFL is formulated as:
\begin{equation}
\min _{r \in \mathcal{R}} \frac{1}{n} \sum_{i=1}^n \min _{h_i \in \mathcal{H}} f_i\left(h_i, r\right),
\end{equation}
where $f_i\left(h_i, r\right)$ is an expected risk function.
In this way, clients can collaboratively learn shared $m_{r}$ using data from all clients while optimizing private $m_{h_{i}}$ only using their own personal data.

\item \textbf{Embedding task} is to achieve watermark embedding in PFL. 
Given a model $m$ consisting of $m_{r}$ and $m_{h_{i}}$, the embedding task is to embed an $l$-bit vector $b \in \{0,1\}^{l}$ into the parameters of one or more layers of the personalized model. 
Similar to prior works~\cite{uchida_embedding_2017,li_fedipr_2022}, we introduce the regularization term to the learning objective of the main task and use the Hamming distance $H\left(b,\tilde{b}\right)$ to verify the embedding effect of the watermark, where $b$ is the target watermark to be embedded and $\tilde{b}$ is the extracted watermark.
After watermark embedding, each client's personalized model should include a private watermark that cannot be disrupted by other watermarks and resists the proposed tampering attack and other common watermarking attacks, thus distinguishing it from previous solutions.
\end{itemize}


\section{Proposed Scheme}\label{sec:Proposed Scheme}


\begin{figure*}[t]
	\centering	\includegraphics[width=0.95\textwidth]{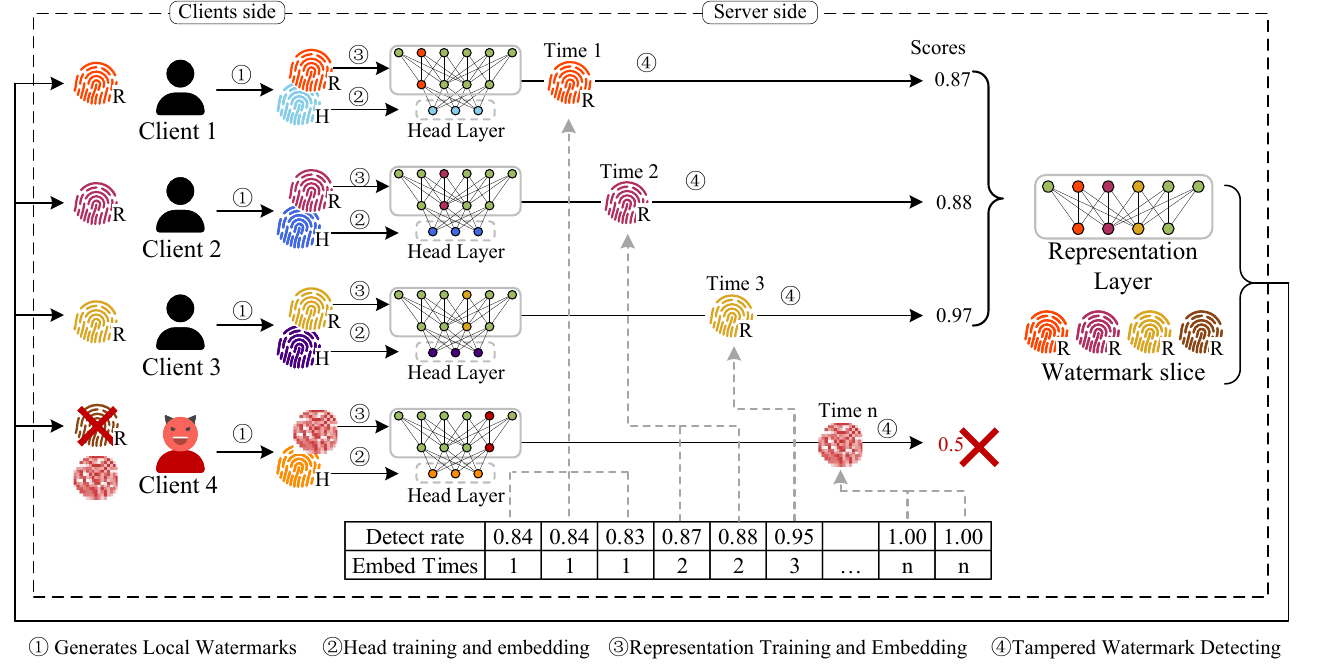}
	\caption{An illustration of RobWE.}
	\label{figs:illustration}
 \vspace{-1em} 
\end{figure*}

\subsection{Overview of RobWE}
The RobWE framework is shown in ~\cref{figs:illustration}, and the workflow in each step is described as follows.
\begin{itemize}

    \item \textbf{System Setup.} Each client $i$ generates local watermarks, including a private watermark information $\Theta_i$ and a watermark slice information $\Omega_i$ obtained from the server.
    
    \item \textbf{Watermark Decoupled Embedding.} Clients embed watermark slices and private watermarks in the representation layer $m_r$ and model head layer $m_h$, respectively. 
    
    \item \textbf{Tampered Watermark Detection.} The server detects and rejects clients who attempt to embed tampered watermarks before each round of model aggregation.
\end{itemize}

\subsection{System Setup}\label{System Setup}
Each client $i$ firstly obtains the initial global model $m^{0}$ from the server and generates its own private watermark information $\Theta=\left(b_i, \theta_i, E_i\right)$, where $b_i$ is the watermark to be embedded into $m_{h_i}$, $\theta_i$ denotes the location of the watermark, and $E_i$ denotes the embedding matrix. 
Notably, the size of $E_i$ varies according to $\theta_i$. 
For example, if client $i$ sets $\theta_i=\{c_1,c_2\}$, this indicates that $b_i$ is to be embedded into the first and second convolutional layers $c_1$ and $c_2$. 
Then, $b_i$ will be divided into $b_{i}^{c_1}$ and $b_{i}^{c_2}$ according to~\cref{eq:b_i}, where $l$ is the length of $b_i$ and the subscript $j$ represents the index of each bit in $b_i$. 
The symbol $||$ denotes the number of parameters in the corresponding layer.
Accordingly, client $i$ generates embedding matrices $E_i=\{E_i^{c_{1}},E_i^{c_{2}}\}$, where $E_i^{c_k} (k\in\{1,2\})$ is generated from the standard normal distribution $\mathcal{N}\left(0,1\right)$ and the size is $|c_k|\times|b_i^{c_k}|$.

\begin{equation}
\left\{\begin{matrix}\begin{aligned}
b_{i}^{c_1} &= \{b_{ij}\}_{j = 1}^{j = \left \lfloor \frac{|c_1|}{|c_1|+|c_2|}\times l \right \rfloor}   
\\
b_{i}^{c_2} &= \{b_{ij}\}_{j = \left \lfloor \frac{|c_1|}{|c_1|+|c_2|}\times l \right \rfloor+1}^{j = l}. 
  \end{aligned}
\end{matrix}\right.
\label{eq:b_i}
\end{equation}

In addition, the server generates a common watermark $b_S$ and divides it into various watermark slices $b_{S_i}$. 
The $b_{S_i}$ is then sent privately to client $i$ along with the embedding location $\omega_{S_i}$ and embedding matrix $E_{S_{i}}$, i.e., the watermark slice information $\Omega_{i}$ equals $(b_{S_i},\omega_{S_i},E_{S_i})$. 
Since $b_{S_i}$ will be embedded in the common part of the model, to avoid watermarking conflicts, $\omega_{S_i}$ requires not to overlap each other.

\subsection{Watermark Decoupled Embedding}\label{Watermark Decoupled Embedding}
In this step, clients should complete the main and embedding tasks. The main idea of this step is to decouple the watermark embedding and model training. Specifically, model training involves head training and representation training, while model aggregation only occurs during representation training. Details of these operations are described below and summarized in~\cref{alg:WDE}.

\begin{algorithm}[t]
    \caption{Watermark Decoupled Embedding} 
    \label{alg:WDE} 
    \begin{algorithmic}[1]
        \REQUIRE $p$ : sampling rate; $n$: number of clients; $T$ :  total training rounds; $\Omega_i$ : watermark slice information; $\lambda$ : learning rate; $m^0$: the initial global model;
        \ENSURE Personalized models $m_i$ with watermarks
        \STATE Server sends $\Omega_i$ and $m^0$ to each client $i$  
        \FOR{ $t=1,2,...,T$}
            \STATE Server samples a batch of clients $\mathcal{C}^{t}$ of size $pn$
            \STATE Server sends representation model $m_{r}^{t}$  to clients
            \FOR{client $i$ in $\mathcal{C}^{t}$}
                \STATE Client $i$ initializes $m_{h_{i}}^{t} \gets m_{h_{i}}^{t-1,\xi}$
                \FOR{$\epsilon=1$ to $\xi$}
                \item 
               $m_{h_{i}}^{t,\epsilon+1}\leftarrow\mathrm{GRD}(L_{h_i}, m_{h_{i}}^{t,\epsilon},\lambda)$
                \ENDFOR
                \STATE $m_{r_{i}}^{t+1}\leftarrow\mathrm{GRD}(L_{r_i}, m_{r_{i}}^{t, \xi},\lambda)$
                \STATE Client $i$ sends updated $m_{r_{i}}^{t+1}$ to server
            \ENDFOR
            \FOR{client $i$ not in $\mathcal{C}^{t}$}
                \STATE Set $m_{h_{i}}^{t+1,\xi} \gets m_{h_{i}}^{t,\xi}$
            \ENDFOR
            \STATE Server aggregates the new representation as
            \STATE $m_{r}^{t+1}=\frac{1}{pn}\sum_{i\in\mathcal{C}^{t}}m_{r_i}^{t+1}$
        \ENDFOR
    \RETURN $m_{i}$ consists of $m_{h_i}^{T,\xi}$ and $m_{r}^{T}$
    \end{algorithmic} 
\end{algorithm}

\begin{itemize}[leftmargin=*]

\item \textbf{Head Training and Embedding.} 
In each round of model training, a certain percentage $p\in (0,1]$ of clients are sampled to participate in the workflow. 
For each selected client $i$, it performs $\xi$ times local update and watermark embedding only in the head layer of the model. 
Specifically, in the $t$ training round, for $\epsilon=1,2,\dots,\xi$, client $i$ adopts the embedding loss defined in~\cref{eq:loss_hi} to update $m_{h_i}^{t,\epsilon}$, i.e., the head layer of model,
\begin{equation}
L_{h_i} = L_{\mathcal{D}_{i}}(m^{t,\epsilon})+L_{\Theta_i}(m_{h_i}^{t,\epsilon}),
\label{eq:loss_hi}
\end{equation}
where $L_{\mathcal{D}_{i}}(m^{t,\epsilon})$ and $L_{\Theta_i}(m_{h_i}^{t,\epsilon})$ are the loss function of the main task and embedding task for embedding a private watermark in $m_{h_{i}}$.
Followed by prior works~\cite{li_fedipr_2022,uchida_embedding_2017}, $L_{\Theta_i}(m_{h_i}^{t,\epsilon})$ can be defined as~\cref{eq:loss_theta},
\begin{equation}
L_{\Theta_i}(m_{h_i}^{t,\epsilon}) = BCE(b_{i},\tilde{b}_i )=BCE(E_i\times \overline{m_{h_i}^{t,\epsilon}},b_{i}),
\label{eq:loss_theta}
\end{equation}
where $B\!C\!E(\cdot)$ is the binary cross entropy function and $\overline{m_{h_i}^{t,\epsilon}}$ is the one-dimensional vector after flattening $m_{h_i}^{t,\epsilon}$. 
Formally, the head training and embedding can be expressed as~\cref{eq:hte}, 
\begin{equation}
m_{h_{i}}^{t,\epsilon+1}\leftarrow\mathrm{GRD}(L_{h_i}, m_{h_{i}}^{t,\epsilon},\lambda),
\label{eq:hte}
\end{equation}
where $G\!R\!D(\cdot)$ represents a gradient updating method such as stochastic gradient descent (SGD)~\cite{Li2022SGD}, and $\lambda$ represents the learning rate.

\item \textbf{Representation Training and Embedding.}
After performing head training and embedding, the sampled clients perform representation training and embedding once only in $m_{r_{i}}^{t}$, i.e., the representation layer of the model. Specifically, the embedding loss for updating $m_{r_{i}}^{t}$ is denoted as~\cref{eq:loss_r},
\begin{equation}
L_{r_{i}} = L_{\mathcal{D}_{i}}(m^{t,\xi})+L_{\Omega_i}(m_{r_{i}}^{t,\xi}),
\label{eq:loss_r}
\end{equation}
where $L_{\Omega_i}(m_{r_{i}}^{t,\xi})$ is defined in~\cref{eq:loss_omega}, which represents the embedding loss caused by the inconsistency between the watermark slice extracted from the $m_{r_{i}}^{t}$ (i.e., $\tilde{b}_{S_i}$) and the target watermark slice (i.e., $b_{S_{i}}$).
\begin{equation}
\begin{aligned}
L_{\Omega_i}(m_{r_{i}}^{t,\xi}) &= BCE(b_{S_i},\tilde{b}_{S_i} ) \\
  &=BCE(E_{S_i}\times \overline{m_{r_{i}}^{t,\xi}(\omega_{S_i})},b_{S_i}).
\end{aligned} 
\label{eq:loss_omega}
\end{equation}

Notably, the symbol $\overline{m_{r_{i}}^{t, \xi}(\omega_{S_i})}$ represents only the parameter at the $\omega_{S_i}$ position of $m_{r_{i}}^{t, \xi}$ that needs to be flattened since $b_{S_i}$ is only embedded in certain regions of the representation layer.
Likewise, the representation training and embedding can be formalized as in~\cref{eq:rte}.
\begin{equation}
m_{r_{i}}^{t+1}\leftarrow\mathrm{GRD}(L_{r_i}, m_{r_{i}}^{t, \xi},\lambda).
\label{eq:rte}
\end{equation}

Once the model training and embedding described above are completed, each sampled client only sends $m_{r_{i}}^{t+1}$ to the server.
If we ignore the presence of malicious clients in the FL scenario, the server can aggregate the representation layer models uploaded by the sampled clients according to~\cref{eq:nomal_avg}, where $\mathcal{C}^{t}$ contains indexes of the sampled client in the $t$ round.
\begin{equation}
m_{r}^{t+1}=\frac{1}{pn}\sum_{i\in\mathcal{C}^{t}}m_{r_i}^{t+1}.
\label{eq:nomal_avg}
\end{equation}
\end{itemize}

\subsection{Tampered Watermark Detection}\label{Tampered Watermark Detecting}
Due to the risk of malicious clients embedding tampered watermarks in FL scenarios,  we designed a tampered watermark detection mechanism. 
As observed from some experiments, we find that the watermark detection rates of malicious clients implementing tampered watermark embedding, along with the watermark detection rates of honest clients embedding watermarks normally, each basically obeys a normal distribution. 
Some results of the experiments can be seen in~\cref{figs:qq}.
Unfortunately, client sampling results in varying embedding times of the watermark slice, and the watermark detection rate varies significantly between clients in the same round, making detection very difficult.

To solve the above issues, the server first records watermark slice detection rates and embedding times for each round of the uploaded $m_{r_i}^{t+1}$. 
Then, to avoid the inability to directly compare watermark detection rates horizontally due to sampling, for each $m_{r_i}^{t+1}$, the server searches all watermark slice detection rate records with the same embedding times as $m_{r_i}^{t+1}$. 
Finally, based on the three-sigma rule commonly adopted in anomaly detection schemes~\cite{Ohana_Anomaly_Detection2022,Liu_Gaussian_2022}, we design the following decision rule to detect malicious clients.
\begin{enumerate}[]
    \item During the early stages of training, when the server has not detected enough malicious clients  (i.e., the number of detected malicious clients satisfies $num_{m} <\beta$), the watermark slice detection accuracy $acc_{i}$ defined in~\cref{eq:acc} should satisfy~\cref{eq:nomal_detect}.
\begin{equation}
acc_{i} = 1 - \frac{1}{|b_{S_i}|} H(b_{S_i},\tilde{b}_{S_i}) ,
\label{eq:acc}
\end{equation}

\begin{equation}
acc_{i} > \mu_{n} -  Z_{C_{n}} \times \frac{\sigma_{n}}{\sqrt{num_{n}} },
\label{eq:nomal_detect}
\end{equation}
    where $\mu_{n}$ and $\sigma_{n}$ are the mean and standard deviation of the detection rate of the searched watermarked slice with the same embedding times as $m_{r_i}^{t+1}$, respectively.
    The number of searched watermarked slices is denoted as $num_n$, and $Z_{C_{n}}$ is the critical value that can be determined by confidence level $C_n$. 
    In practice, $C_n$ can be determined on a practical basis. For example, if tolerance for malicious clients is high, $C_n$ can be set larger so that most of the clients will not be considered malicious early in the training.

    \item When the number of malicious clients detected by the server is sufficient to characterize malicious clients, i.e., $(num_m \geq \beta)$, $acc_{i}$ should satisfy~\cref{eq:maliciou_detect}, which implies $acc_i$ does not match the characterize of malicious clients.
    \begin{equation}
acc_{i} > \mu_{m} +  Z_{C_{m}} \times \frac{\sigma_{m}}{\sqrt{num_{m}} } ,
\label{eq:maliciou_detect}
\end{equation}
    where $\mu_{m}$ and $\sigma_m$ are the mean and standard deviation of the $acc_i$ of the detected malicious clients, respectively.
    And $Z_{C_{m}}$ is the critical value that can be determined by confidence level $C_m$. The $C_m$ represents the optimism about accepting honest clients. For example, setting $C_m$ to a relatively small value can ensure that most clients meet the upper bound requirement for the distribution of watermark detection rate among malicious clients, provided we assume that most clients are honest.
\end{enumerate}
\begin{table*}
\centering
\begin{tabular}{c|c|cccc|c} 
\hline
\multirow{2}{*}{DataSet}  & \multirow{2}{*}{NonIID} & \multicolumn{5}{c}{RobWE (v.s. FedIPR)}                                                                          \\ 
\cline{3-7}
                          &                         & 0 bit         & 50 bits          & 100 bits         & 150 bits          & Gap                                      \\ 
\hline
\multirow{6}{*}{CIFAR-10} & Dir(0.1)                & 88.11 (26.63) & 87.77 (15.09) & 87.99 (12.88) & 87.53 (14.40) & \textbf{0.66\%}~(45.93\%)  \\
                          & Dir(0.3)                & 75.70 (46.94) & 74.89 (33.65) & 73.76 (21.40) & 73.63 (12.77) & \textbf{2.72\%}~(72.80\%)  \\
                          & Dir(0.5)                & 68.82 (68.28) & 68.69 (29.01) & 68.09 (11.74) & 67.71 (9.98)  & \textbf{1.61\%}~(85.38\%)  \\
                          & K(2)                    & 87.48 (17.31) & 86.62 (15.56) & 86.18 (12.46) & 86.63 (12.32) & \textbf{0.96\%}~(28.83\%)           \\
                          & K(4)                    & 70.43 (62.07) & 70.57 (19.22) & 69.31 (17.04) & 66.25 (13.24) & \textbf{5.93}\%~(78.67\%)           \\
                          & K(6)                    & 64.31 (69.73) & 62.13 (38.54) & 61.19 (21.11) & 58.61 (14.26) & \textbf{8.85\%}~(79.55\%)         \\ 
\hline
\multirow{6}{*}{MNIST}    & Dir(0.1)                & 99.38 (95.80) & 99.36 (81.13) & 99.28 (50.23) & 99.23 (22.98) & \textbf{0.14\%}~(76.01\%)  \\
                          & Dir(0.3)                & 98.44 (97.72) & 98.35 (94.92) & 98.30 (50.24) & 98.30 (11.35) & \textbf{0.14\%}~(88.39\%)  \\
                          & Dir(0.5)                & 97.96 (98.21) & 97.76 (92.75) & 97.49 (76.56) & 97.44 (56.83) & \textbf{0.54\%}~(42.13\%)  \\
                          & K(2)                    & 99.37 (97.29) & 99.31 (23.06) & 99.20 (16.49) & 98.54 (11.03) & \textbf{0.84\%}~(88.67\%)  \\
                          & K(4)                    & 97.90 (97.11) & 97.85 (95.62) & 97.54 (60.46) & 97.37 (16.03) & \textbf{0.53\%}~(83.49\%)  \\
                          & K(6)                    & 88.64 (98.54) & 88.12 (95.11) & 87.15 (81.08) & 87.00 (38.00) & \textbf{1.84\%}~(61.44\%)  \\
\hline
\multirow{6}{*}{FMNIST}   & Dir(0.1)                & 96.89 (64.04) & 96.80 (38.85) & 96.70 (20.50) & 96.74 (11.02) & \textbf{0.15\%}~(82.79\%)  \\
                          & Dir(0.3)                & 91.79 (84.91) & 91.64 (80.74) & 91.10 (66.89) & 89.73 (39.12) & \textbf{2.24\%}~(53.93\%)  \\
                          & Dir(0.5)                & 90.94 (86.76) & 90.92 (82.60) & 90.13 (69.43) & 89.92 (54.20)  & \textbf{1.12\%}~(37.53\%)  \\
                          & K(2)                    & 95.99 (80.07) & 95.02 (62.47) & 92.00 (65.59) & 91.08 (51.93) & \textbf{5.12\%}~(35.14\%)           \\
                          & K(4)                    & 92.60 (81.51) & 92.44 (67.72) & 92.21 (37.28) & 92.05 (16.66) & \textbf{0.59\%}~(79.56\%)           \\
                          & K(6)                    & 79.43 (85.75) & 78.10 (82.42) & 78.09 (70.17) & 76.95 (24.30) & \textbf{3.12\%}~(71.66\%)         \\ 
\hline
\end{tabular}
\caption {
The model accuracies after embedding different bits of watermarks and the relative variation values (i.e., Gap) of the model accuracies when embedding the maximum watermarks compared to those when the watermarks are not embedded.
}
\label{tab:fidelity}
\vspace{-1em} 
\end{table*}

\section{Experiments}\label{experi}
\subsection{Experimental Setup}

\begin{itemize}[leftmargin=*]
\item\textbf{Datasets.}
We adopt CIFAR-10~\cite{krizhevsky2009learning}, MNIST~\cite{lecun1998gradient}  and FMNIST~\cite{xiao2017/online} datasets, and utilize the Non-IID setup ~\cite{noniid_2022} for dataset division.
Here, the notation $K(n)$ stands for assigning $n$ differently labeled data to each client, and $Dir(\beta)$ stands for constructing the data distribution for each client according to the Dirichlet distribution.
\item\textbf{Baseline.}
We compare RobWE with FedIPR~\cite{li_fedipr_2022}, which allows clients to embed both backdoor-based and feature-based watermarks in the global model for ownership protection. To ensure a fair comparison, we solely use feature-based watermarks in FedIPR.

\item\textbf{Models and Configurations.}
We adopt convolutional neural network (CNN)~\cite{collinsExploitingSharedRepresentations2021} as the baseline model for PFL training, facilitating the verification of schemes’ ability to embed watermarks on models with fewer parameters.
We employ 100 clients using the SGD optimizer with a learning rate of 0.01 over 50 epochs. Local training consists of 11 epochs, divided into 10 for the head layer training and 1 for the representation layer training. The batch size is 10 with a sampling rate of 0.1.

\item \textbf{Metrics.}
We evaluate RobWE in the following aspects:
\begin{enumerate}[]
    \item \textbf{Fidelity}.
     We adopt the accuracy of the main task as the fidelity metric and investigate the relative variation in accuracy when adding the maximum watermarked bits, defined as Gap, compared to a model without any embedded watermarks. A smaller Gap indicates higher fidelity.
    \item \textbf{Reliability}. We utilize the watermark detection rate and the range of variation under embedding different sized watermarks to measure reliability.
    \item \textbf{Robustness}. We use watermark detection rate $w$ to reflect robustness under pruning, fine-tuning, and tampering attacks. For tampering attacks, the tamper   detection rate $d_{t}$, false positive rate $d_{f}$  and the discrepancy between watermark detection rates of honest and malicious clients, denoted as $\Delta _ {w_{n}, w_{m}}$, are employed to gauge the defensive effectiveness.
 
\end{enumerate}
\end{itemize}


\subsection{Experimental Results}
\subsubsection{Fidelity Performance}
We embed watermarks with various bits using RobWE and FedIPR under different Non-IID settings. \cref{tab:fidelity} shows that RobWE has a higher model accuracy compared to FedIPR in most cases, with Gap values ranging from 0.14\% to 8.85\% much smaller than the range of Gap values for FedIPR. 
This indicates that RobWE is not only able to obtain personalized models with better accuracy, but also has very little negative impact on model accuracy as the number of watermark embedded bits increases.
On the contrary, for FedIPR, the accuracy steadily decreases as the number of watermarked bits increases, with a Gap between 28.83\% and 88.67\%.
The model accuracy in the worst scenario drops to 11.03\%, a drop of 88.67\% compared to the case without watermarking, making the model unadoptable.

\subsubsection{Reliability Performance}
\begin{figure}[t]
   \centering
   \includegraphics[width=\linewidth]{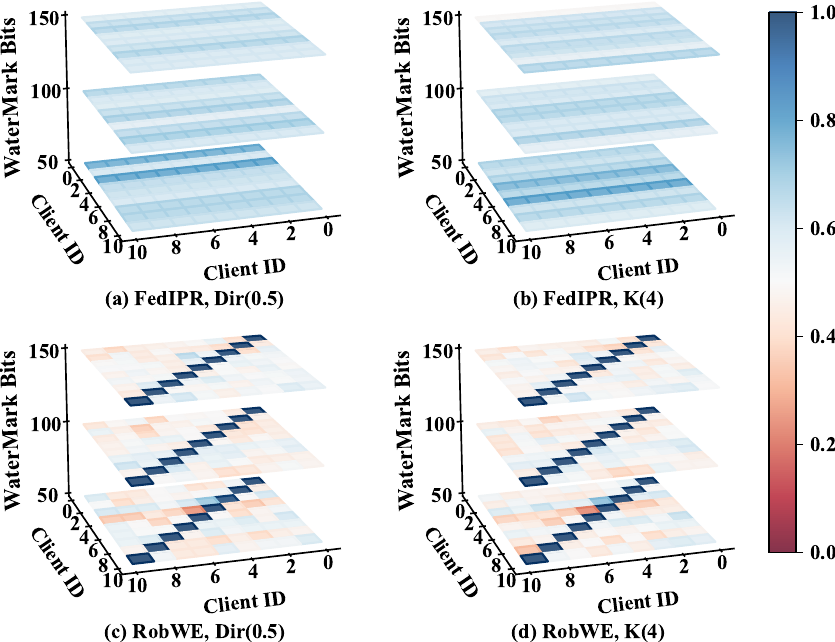}\\
   \caption{The watermark detection rate for other client watermarks on each client' model (non-diagonal region) and the detection rate for own private watermarks (diagonal region).}
   \label{figs:heatmap}
\end{figure}

Following~\cite{li_fedipr_2022}, we select 10 clients to participate in model watermark embedding without sampling operation. Each client embeds a private watermark of 50 bits, 100 bits and 150 bits under $Dir(0.5)$ and $K(4)$ Non-IID settings, respectively. After embedding, we extract the watermarks of all clients on each client's model and calculate the watermark detection rate.
As shown in~\cref{figs:heatmap}, we use multiple heatmaps to show the detection rate of all clients' watermarks detected on each client's private model, with darker colors representing higher detection rates.
The diagonal position represents the detection rate of the client's own private watermark.
The results show that the watermark detection rate of each client's private watermark in RobWE is significantly greater than that of other clients (non-diagonal region), while the private watermark detection rate of each client in FedIPR is even worse than that of other clients' watermarks since the client's model is mixed with the watermarks of other clients.
Additionally, the baseline model we used for embedding watermarks has only 128 channels.
As shown in~\cref{tab:reliability}, RobWE also significantly outperforms FedIPR when the watermark occupancy is much larger than the maximum watermark occupancy threshold (10\%, where FedIPR can guarantee a watermark detection rate of 1).

\begin{table}[t]
\centering
\begin{tabular}{c|c|c} 
\hline
\multicolumn{1}{l|}{Bits} & \multicolumn{1}{l|}{Occupancy Ratio} & \multicolumn{1}{l}{Range (RobWE vs. FedIPR)}  \\ 
\hline
50                        & 39.06\%                         & (0.98,1)/(0.58,0.78)                        \\ 
\hline
100                       & 78.13\%                         & (1,1)/(0.57,0.68)                        \\ 
\hline
150                       & 117\%                           & (0.99,1)/(0.49,0.66)                     \\
\hline
\end{tabular}
\caption{The model channel occupancy ratio for different size of watermarks and the range of variation in the watermark detection rate under corresponding embedding scenarios.}
\label{tab:reliability}
\end{table}

\subsubsection{Robustness Performance}
\begin{figure}[t]
   \centering
   \includegraphics[width=\linewidth]{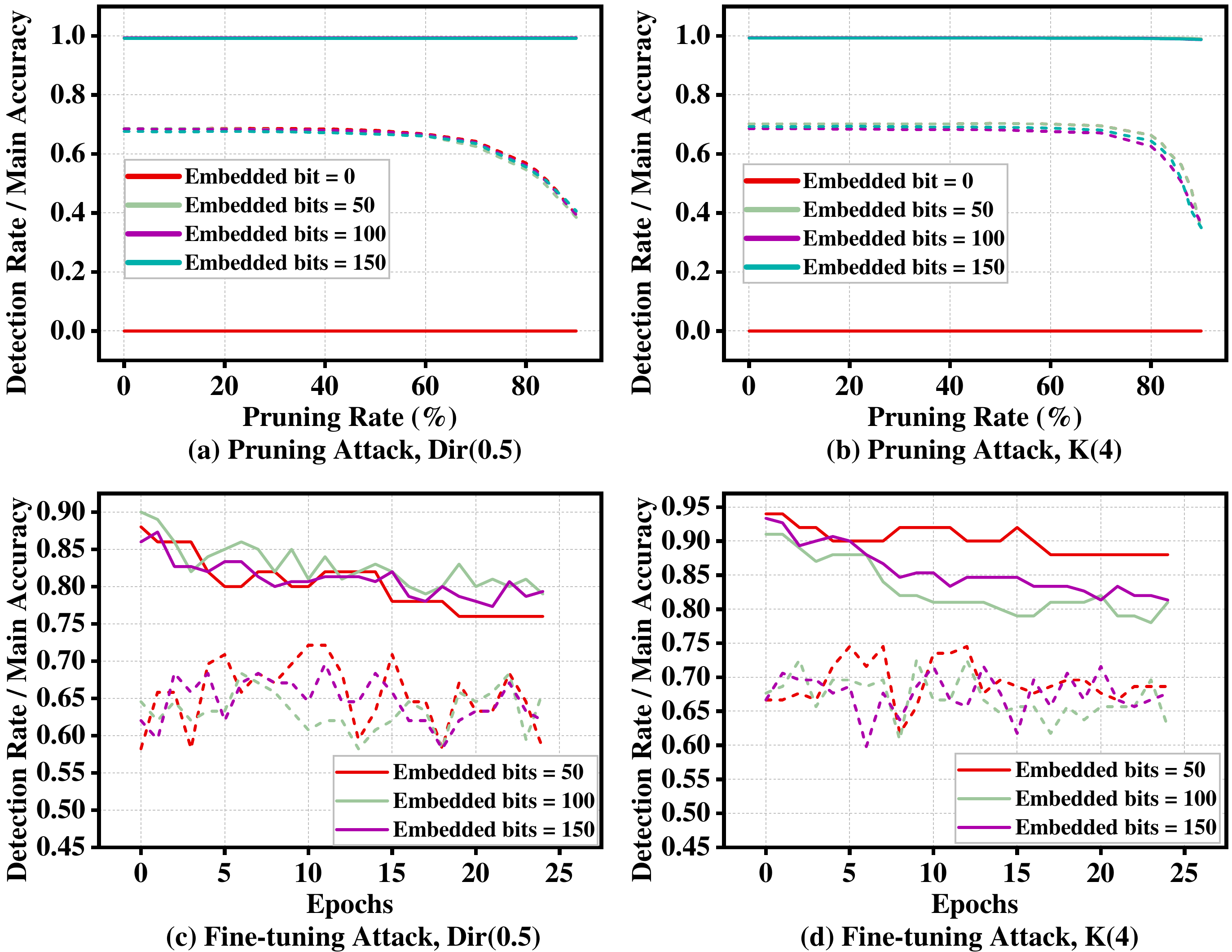}\\
   \caption{In the Non-IID setting, the model accuracy (dashed line) and watermark detection rates (solid line) of models with different embedded bits watermark after attacks.}
   \label{figs:attack}
\end{figure}

To illustrate robustness to common attacks on watermarking, we perform a pruning attack~\cite{see2016compression} and a fine-tuning attack~\cite{li_fedipr_2022} against private watermarks embedded in the head layer, respectively.
As shown in~\cref{figs:attack}(a) and (b), as the pruning rate increases, more and more model parameters are removed, and when the ratio reaches 70\%, the accuracy of the model already starts to decrease, but the watermark detection rate remains stable at a high level.
Therefore, RobWE is highly resistant to pruning attacks.

To implement fine-tuning attacks, honest clients first train 50 rounds and embed their private watermarks. Then, the attacker locally trains 25 rounds of fine-tuning based on these trained models.
\cref{figs:attack}(c) and (d) show that although the watermark detection rate decreases during the model fine-tuning process, the overall rate remains above 0.75, which is still effective for ownership authentication.

Then, we perform the adaptive tampering attack against the representation layer of the model.
The effect has been displayed in~\cref{figs:tampered-attack}, where malicious clients' watermark detection rate is greater than or obfuscates honest clients' watermarks.
We further increase the fraction of malicious clients $f_m$ and the watermark tampering rate $f_t$. The larger the gap between the watermark detection rate of honest clients and the watermark detection rate of malicious clients $\Delta _ {w_{n}, w_{m}}$ after applying defenses indicates greater resistance to tampering attacks.
\cref{tab: detection} shows that $\Delta _ {w_{n}, w_{m}}$ is still significant even if the malicious client has only been tampered with a very small portion (i.e., 0.1). In fact, the smaller the $f_t$, the harder it is for the attacker to be detected.
Under various settings, our method consistently maintains a malicious client detection rate of more than 0.9 and a maximum false positive rate of no more than 0.05.

\begin{figure}[t]
   \centering
   \includegraphics[width=\linewidth]{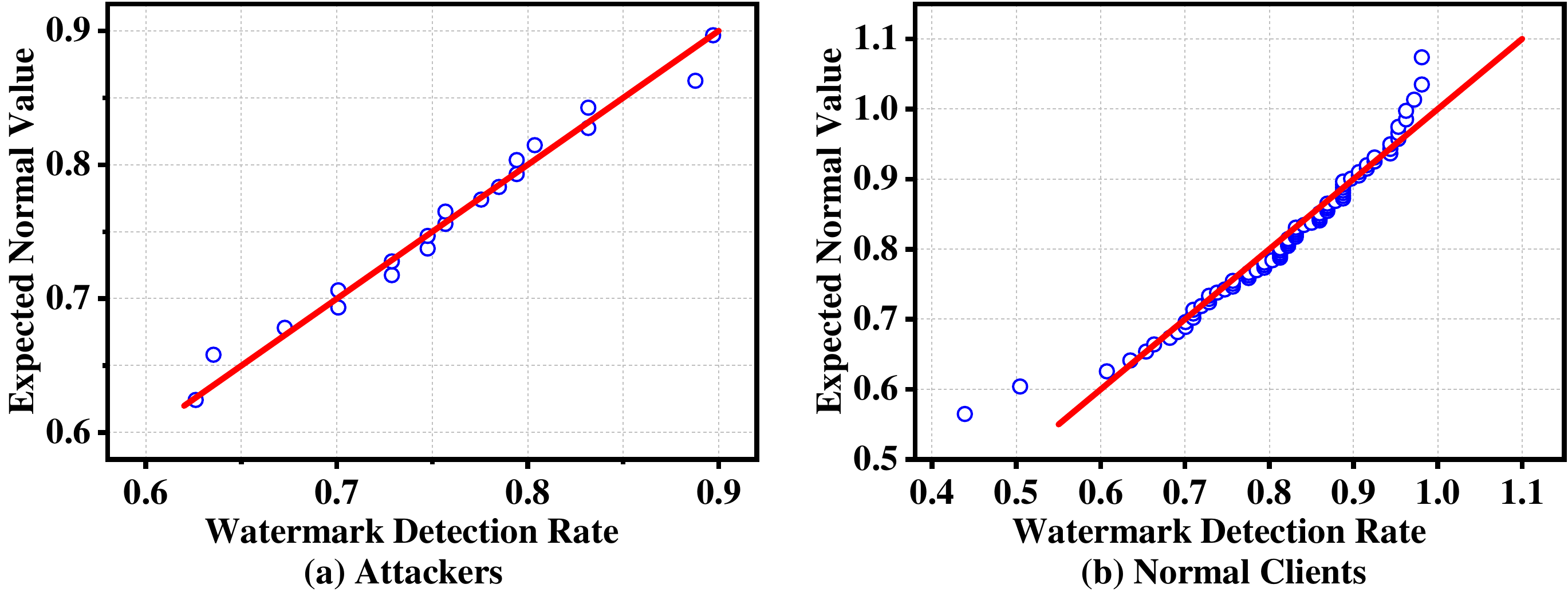}\\
  \caption{The Quantile-Quantile plot for assessing  whether the watermark detection rate between normal clients and attackers obeys a normal distribution.}
  \label{figs:qq}
\end{figure}

Meanwhile, \cref{figs:qq} shows that the watermark detection rates of malicious and honest customers basically satisfy the normal distribution.
In fact, we chose $C_n$ of 0.975 and $C_m$ of 0.5 for detection. 
The former is a direct adoption of the $3\sigma$ detection interval, while the latter adopts neutrality to accept honest clients.
In practice, this can be adjusted according to the actual situation, while not being limited to the requirement of strictly meeting the normal distribution.


\begin{table}[t]
\resizebox{\columnwidth}{!}{%
\begin{tabular}{c|c|c|c|c|c|c}
\hline
NonIID                    & ($fm$,$f_{t}$) & $w_{n}$ & $w_{m}$ & $d_{t}$ & $d_{f}$ & $\Delta _ {w_{n}, w_{m}}$   \\ \hline
\multirow{4}{*}{Dir(0.5)} & (0.2, 0.1)     & 95.42   & 86.65   & 0.95    & 0.01    & 8.77  \\
                          & (0.2, 0.3)     & 96.19   & 87.33   & 1.0     & 0       & 8.86  \\
                          & (0.4, 0.1)     & 97.45   & 79.44   & 0.95    & 0       & 18.01 \\
                          & (0.4, 0.3)     & 96.34   & 79.97   & 0.9     & 0       & 16.37 \\ \hline
\multirow{4}{*}{K(4)}     & (0.2, 0.1)     & 95.43   & 84.29   & 1.0     & 0.05    & 11.14 \\
                          & (0.2, 0.3)     & 95.43   & 86.24   & 1.0     & 0.04    & 9.19  \\
                          & (0.4, 0.1)     & 95.93   & 79.11   & 0.975   & 0       & 16.82 \\
                          & (0.4, 0.3)     & 94.61   & 80.03   & 0.925   & 0       & 14.58 \\ \hline
\end{tabular}%
}
\caption{The detection performance ($d_{t}$) of RobWE on malicious clients under various proportions of malicious clients ($f_{m}$) and tampering rates ($f_{t}$).}
\label{tab: detection}
\end{table}

\section{Conclusion}\label{conclude}
This paper presents a robust watermark embedding scheme, named RobWE, to protect the ownership of personalized models in PFL.
RobWE decouples the watermark embedding procedure into head layer embedding and representation layer embedding, allowing clients to embed individual private watermarks independently of model aggregation. Furthermore, we propose a watermark slice embedding method for embedding non-overlapping watermark slices in the shared representation layer to facilitate contribution proof by each client.
Additionally, for the proposed tampering attack, we couple with a dedicated detection scheme aimed at improving robustness.
Through a series of experiments, we comprehensively evaluate the performance of RobWE, and the results show that our scheme successfully safeguards personalized model ownership in PFL. In the future, we will consider designing tamper-resistant watermark embedding schemes under both an untrusted server and malicious clients. 
We also plan to design a model leakage tracing mechanism for PFL.
{
    \small
    \bibliographystyle{ieeenat_fullname}
    \bibliography{main}
}


\end{document}